\documentclass{aa501}
\usepackage{longtable}
\usepackage{lscape}
\usepackage{latexsym}
\usepackage{amssymb}
\usepackage[dvips]{graphicx}

\def\lsol{~L$_\odot$ }

\def\micron{\,$\mu$m}
 
\def\marc{mag~arcsec$^{-2}$}

\begin{document}

\title{1.65~$\rm \mu$m (H-band) surface photometry of galaxies. VII:
dwarf galaxies in the Virgo Cluster.
\thanks{Based on observations 
taken with the ESO/NTT (ESO program 64.N-0288), 
with the Telescopio Nazionale Galileo (TNG) operated
on the island of La Palma by the Centro Galileo Galilei of the CNAA
at the Spanish Observatorio del Roque de los Muchachos of the IAC,
with the San Pedro Martir 2.1~m telescope of the Observatorio 
Astronomico Nacional (OAN, Mexico), and with the
OHP 1.2~m telescope, operated by the French CNRS.}
}

\author{G. Gavazzi\inst{1}
\and S. Zibetti \inst{1}
\and A. Boselli\inst{2}
\and P. Franzetti \inst{1}
\and M. Scodeggio \inst{3}
\and S. Martocchi \inst{1}
}

\offprints{G. Gavazzi}

\institute{Universit\`a degli Studi di Milano - Bicocca, P.zza 
dell'Ateneo Nuovo 1, 20126 Milano, Italy
\and
Institut d'Astrophysique de Marseille, Traverse du Siphon, F-13376 Marseille
Cedex 12, France
\and 
Istituto di Fisica Cosmica ``G. Occhialini'', CNR, via Bassini 15, 
20133, Milano, Italy
}

\date{Received 21 December 2000; accepted 26 March 2001}

\abstract{We present near-infrared H-band (1.65\micron ) observations and
surface brightness profile decompositions for 75 faint ($13.5 \lesssim
m_{\rm p} \lesssim 18.5$) galaxies, primarily taken among dwarf Ellipticals
members of the Virgo cluster, with some Centaurus Cluster members, a
BCD and two peculiar galaxies taken as fillers.  We model their
surface brightness profiles with a de Vaucouleurs (D), exponential
(E), mixed (bulge+disk or M) or truncated (T) law, and we derive for
each galaxy the H band effective surface brightness ($\mu_{\rm e}$) and
effective radius ($r_{\rm e}$), the asymptotic total magnitude $H_{\rm T}$ and the
light concentration index $C_{31}$, defined as the ratio
between the radii that enclose 75\% and 25\% of the total light
$H_{\rm T}$.  For a subsample we compare the NIR surface photometry with
similar data taken in the B and V bands, and we give the B-H and B-V
color profiles.\\ Combining the present data with those previously
obtained by our group (1157 objects) we analyze the NIR properties of
a nearly complete sample, representative of galaxies of all
morphological types, spanning 4 decades in luminosity.  We confirm our
earlier claim that the presence of cusps and extended haloes in the
light profiles ($C_{31}>5$) is a strong, non-linear function of
the total luminosity.  We also find that: i) among dEs and dS0s
galaxies D profiles are absent; 50\% of the decompositions are of type
M, the remaining being of type E or T.  ii) Less than 50\% of the
giant elliptical galaxies have pure D profiles, the majority being
represented by M profiles. iii) Most giant galaxies (from elliptical
to Sb) have M profiles. iv) Most of late type spirals (Scd to BCD)
have either E or T profiles. v) The type of decomposition is a strong
function of the total H band luminosity, independent of the Hubble
classification: the fraction of type E decompositions decreases with
increasing luminosity, while those of type M increase with
luminosity. Pure D profiles are absent in the low luminosity range
$L_{\rm H}<10^{10}$ \lsol and become dominant above $10^{11}$ \lsol, while T
profiles are present only among low luminosity galaxies.  vi) We find
that dE-peculiar galaxies have structural parameters indistinguishable
from those of late-type dwarfs, thus they might represent the missing
link between dEs and dIs.
}
\titlerunning{NIR observations of Virgo galaxies}
\authorrunning{G. Gavazzi, S. Zibetti, A. Boselli et al.}
\maketitle
\section{Introduction}
The comprehension of the physical processes behind the formation and
evolution of galaxies is far from satisfactory, but plausible
scenarios based on hierarchical clustering (Kauffmann \& White 1993,
Kauffmann \& Charlot 1998), developed in the framework of CDM
cosmology, are becoming the current paradigm superseeding previous
models based on the monolithic collapse scenario (Larson 1975, Sandage
1986).\\
The advent of 8m-class telescopes and of the forthcoming NGST will soon 
make it possible to trace observationally the evolution of galaxies up
to redshifts of cosmological relevance, and therefore allow a direct
comparison between model predictions and observations.\\ 
Should however adult galaxies retain some memory of their ``infancy'',
observations carried out at z=0 would provide important constraints on
their evolutionary history.  For these nearby objects even the present
instrumentation is suitable for extending to intrinsically low
luminosity galaxies a detailed determination of their ``shape'' and
``size'' parameters, such as color, morphology, and brightness profile,
or luminosity, radius, and mass, respectively (see Whitmore
1984). These parameters are the constituents of the well known scaling 
relations of local galaxies: the Fundamental Plane for the Ellipticals
(Djorgovski \& Davis 1987, Dressler et al. 1987), and the Tully-Fisher
relation for the Spiral galaxies (Tully \& Fisher 1977).\\
Over the last two decades we have witnessed an extensive effort in
studying these properties, limited however mostly to the optical
bands.  Less systematic efforts were devoted to Near infrared (NIR)
investigations, in spite of these being the most suitable ones for
studying the properties of galaxies, because most of a galaxy luminous
mass sits in the old stellar population traced by NIR light (Gavazzi
et al. 1996c), and because of the greatly reduced dust obscuration at
these wavelengths. To fully exploit these two advantages we have
made extensive use of NIR panoramic detectors to obtain H (and K')
band images of nearby galaxies.  We first concentrated on disk galaxies 
(see Gavazzi et al. 1996a (Paper I), Gavazzi et al. 1996b (Paper II),
Boselli et al.  2000 (Paper IV) and Boselli et al. 1997 (B97)), while
later we extended the survey to the early-types (Gavazzi et al. 2000a
(Paper III)).  Using these data Gavazzi et al.(2000b, Paper V) studied
the structural properties of galaxies that can be derived from
surface-photometry measurements at NIR pass-bands: i.e. their light
profiles.  The observing sample was selected among members of 5
nearby, rich clusters: namely the Virgo, Coma, A1367, A262 and Cancer
clusters, in addition to a significant population of galaxies in the
``Great Wall'', the bridge between Coma and A1367.  The survey
included a representative sample of galaxies spanning all
morphological types (including Im and BCDs), except early-type dwarfs
(dE, dS0) which were severely undersampled because, due to the their
low surface brightness at NIR bandpasses, they could not be observed
with 2m class telescopes.  Beside our work, early panoramic NIR
observations exist for only 15 dwarf elliptical galaxies in the Virgo
cluster, as reported by James (1991, 1994).\\ 
To fill this gap, we have obtained NIR imaging observations of 50
dwarf elliptical and dwarf S0 galaxies, of 11 dI galaxies and of 11
giant galaxies from the Virgo Cluster Catalogue (hereafter VCC,
Binggeli et al. 1985) and from the Centaurus Cluster Catalogue
(hereafter CCC, Jerjen \& Dressler 1997).  From these observations we
derive the azimuthally averaged light profiles, that are fitted using
either a de Vaucouleurs $r^{1/4}$ law, an exponential law, a mixed
(bulge+disk) model, or an exponentially truncated model.  We derive
some relevant photometric parameters, namely: the asymptotic total
magnitude, the effective radius (within which half of the total galaxy
luminosity is enclosed), the effective mean surface brightness, the
light concentration index $C_{31}$, and the bulge to total flux
ratio for the two-component models.  Moreover, using observations
taken in optical bands, we compare light profiles in B, V and H, and
present color profiles.  Combining the present results with those
given in Paper V, we obtain a nearly complete sample, covering all
morphological types and spanning 4 decades in luminosity.\\ The paper
is organized as follows: the sample selection criteria are discussed
in Sect. 2.  The observations and data reduction are described in
Sect.  3.  The procedures adopted to derive the light profiles and
their fitted models are given in Sect. 4.  The results of the present
work are given in Sect. 5. Some implications of the present analysis
on the structural properties of galaxies are discussed in Sect. 6 and
summarized in Sect. 7.

\section{Sample selection}

\begin{table}[!t]
\caption{Sample completeness}
\label{completeness}
\[
\begin{array}{p{0.20\linewidth}ccccccccc}
\hline
\noalign{\smallskip}
&\multicolumn{2}{c}{m_{\rm p} < 14}&
\multicolumn{2}{c}{m_{\rm p} < 15}&
\multicolumn{2}{c}{m_{\rm p} < 16}&
\multicolumn{2}{c}{m_{\rm p} < 18}\\
   & N_{\rm obs} & \%	  & N_{\rm obs} & \% & N_{\rm obs} & \% & N_{\rm obs} & \%\\  
\hline
\noalign{\smallskip}
Coma S.C.      & 21 &(96) & 176 &(99)& 503&(98) &   -  &	- \\
Virgo (ISO) & 96 &(98) & 143 &(83)& 170&(67) &   188&(42)\\
Virgo    & 195&(94) & 260 &(72)& 295&(55) &   315&(34)\\
\hline
\noalign{\smallskip}
\end{array}
\]
\end{table}

We report H-band (1.65\micron ) observations of 52 faint Virgo
galaxies (39 early and 13 late) selected from the VCC catalogue
(Binggeli et al.  1985) (restricted to m$\rm _p \lesssim$~16.0).  With
the aim of achieving the highest completeness level, the highest
priority was given to objects lying in the region either within 2
degrees of projected radial distance from M87 or in the corona between
4 and 6 degrees selected by the ISO consortium for complete MIR and
FIR observations with ISO (see B97 for a detailed description of this
selection criterium).  During periods when Virgo was at airmass
exceeding 2.2, we observed 20 (17 early and 3 late) galaxies in the
Centaurus cluster selected from the CCC as fillers.  We also observed
two Pec galaxies (CGCG 97-073, 97-087) and the BCD galaxy IZw018.\\
Adding the present observations to those available from our previous
Papers I, II, III, IV and B97 of this series, and limiting our sample
to the Coma supercluster region ($\rm 0.0^o \le \delta \le 20.0^o$;
$\rm 11^h30^m \le \alpha \le 13^h30^m$) and to the Virgo cluster, the
complete subsample with NIR observations is as given in Tab.\ref{completeness} in 4
bins of $m_{\rm p}$.  The Coma region is completed at the limit of the
Zwicky Catalogue.  Assuming a distance modulus of Coma of 34.9 ($\rm
H_{\rm 0}=75~km~sec^{-1}~Mpc^{-1}$) this corresponds to $M_{\rm p}=-19.2$,
i.e. the observations cover a complete sample of giant galaxies.
At the distance to the Virgo cluster A of 17 Mpc (Gavazzi et al,
1999), the limiting magnitude $m_{\rm p}=16$ reached by the present
observations corresponds to $M_{\rm p}=-15.2$, i.e. 4 mag deeper than in
Coma.  However the observations of Virgo can be considered complete
down to $M_{\rm p}=-17$, while they are still $\sim 50\%$ complete at
$M_{\rm p}=-15$ (slightly better in the ISO subsample).  The analysis
carried out in Sect. 6 comprises all objects in Tab.\ref{logbooktab}.

\section{Observations and data reduction}
\subsection{The observations}

The NIR observations reported in this paper were acquired in the
photometric nights of March 26th and 27th, 2000 with the 3.6~m, f/11
ESO-NTT telescope, and of December 18th and 19th, 1999 during the
science verification period at the 3.6~m, f/11 TNG telescope.  The
ESO-NTT Nasmyth focus was equipped with the SOFI $1024^2$ pixel array
camera.  With a pixel scale of 0.29 arcsec/pixel, SOFI has a field-of-view
of $\sim5\times5~\rm arcmin$.  The TNG Nasmyth focus was equipped with the
NICMOS3 $256^2$ pixel array camera ARNICA (Lisi et al. 1993; Lisi et al.
1996; Hunt et al. 1996), which, with a pixel scale of 0.352
arcsec/pixel, gives a field-of-view of $\sim1.5\times1.5~\rm arcmin$.\\ The
seeing at ESO-NTT was always sub-arcsecond, except for one observation
taken at a very high airmass (2.14). The mean seeing was 0.77 arcsec
(FWHM), with a minimum of 0.59 arcsec. At TNG we observed with a mean
seeing of 1.5 arcsec.\\
The NIR sky is extremely bright ($\sim 13~\rm mag~arcsec^{-2}$) compared with
the targets ($\sim 22~\rm mag~arcsec^{-2}$), with significant fluctuations (up
to $10\div20\%$ of the mean value) on time-scales comparable with the
duration of one observation. In order to observe in linear and
background limited regime, observations must be split into several elementary
exposures (``coadds'') which are averaged together. In order
to monitor the sky fluctuations, on-target observations are alternated
with off-target observations, following typical pointing sequences
(``mosaics'').\\
Depending on the extension of the observed sources, we used two types of
mosaic\footnote{Sketches of typical ``mosaics'' can be found in Fig. 2 of
B97.}.  If the apparent size of the source is similar to the available
field of view, half of the observing time was spent on the target, half
on the sky.  The 8 on-target positions were chosen with slight offsets
in order to allow for median rejection of bad pixels. These fields were
alternated with 8 sky observations.  This was the case of all TNG
observations and of some NTT fields in which more than one galaxy could
be accomodated into one frame (e.g. 4 Centaurus fields and the galaxy
pair VCC1491-1499).  Thanks to the large field of view of SOFI compared 
to the target objects, most of the NTT observations of objects
$\lesssim 1.5~\rm arcmin$ were performed using a second type of mosaic in
which the target is always in the field, but is moved around in 6 never
overlapping positions, making independent, time costly sky
measurements unnecessary.  We always avoided to set the object in
the north-eastern quadrant of SOFI affected by a lower optical
quality.\\

Tab.\ref{logbooktab} reports the log-book of the observations, including
parameters relevant to NIR observations, as follows:\\
Column (1): VCC (Binggeli et al. 1985), CCC (Jerjen \& Dressler 1997) or CGCG 
(Zwicky et al. 1961-68) denomination.  \newline
Column (2): NGC/IC names. \newline
Column (3), (4): adopted (B1950.0) celestial coordinates, taken from NED 
\footnote{NASA-IPAC Extragalactic Databasa (NED) is operated by the 
Jet Propulsion Laboratory, California Institute of Technology, under 
contract with NASA}, with typically one arcsec uncertainty. \newline
Column (5): the morphological type taken from Binggeli et al. (1985), 
Jerjen \& Dressler (1997).\\
Column (6): the photographic magnitude from the VCC and CCC.\\
Column (7): the observing date.\\
Column (8): the telescope used.\\
Column (9): the number $N_{\rm cds}$ of coadded (averaged) exposures for 
each frame.\\
Column (10): the number $N_{\rm cmb}$ of frames combined to obtain the 
final image.\\
Column (11): the exposure time for each coadded exposure.\\
Column (12): the total integration time.\\
Column (13): the mean airmass during the observation.\\
Column (14): the adopted filter.\\
Column (15): the seeing (FWHM) in arcsecs.\\
\\
Optical photometric observations in the B and V passbands for 24 objects
in the sample were obtained with the San Pedro Martir (SPM) 2.1m Telescope
from April 20 to 24, 1998 (20 galaxies), and with the Observatoire de Haute
Provence (OHP) 120cm telescope from March 1 to 3, 1998 (4 galaxies,
namely VCC608, 745, 1073 and 1254). Both telescopes were equipped with a
TK1024 $1024^2$ pixel CCD camera. The pixel scale is $0.30~\rm arcsec~pix^{-1}$
at SPM, and $0.69~\rm arcsec~pix^{-1}$ at OHP. Exposure times were of 600 sec
for the V-band and of 900 sec for the B-band observations. 

\subsection{Photometric calibration}
Observations of standard stars, from Hunt et al. (1998), and Persson et al.
(1998), listed in Tab.\ref{stdstars}, were taken one per hour
for calibration purposes.
The calibration stars were observed with a third pointing sequence which
consists of five positions, starting with the star near the center of
the array, followed by positioning the star in each of the four
quadrants of the array.  At TNG the telescope was defocussed to avoid
saturation, since we
observed the two brightest stars of the list.  The
typical photometric uncertainty is 0.05~mag, both for ESO-NTT and TNG
observations. 
\begin{table}[!th]
\caption{Standard calibration stars}\label{stdstars}
\begin{tabular}{l|r}
\hline
Star	       &       $H_{\rm mag}$\hfill	     
\\
\hline					     %
AS 08\_0      &	      $8.723\pm0.014$	     
\\
AS 18\_0      &       $12.402\pm0.004$	     
\\
AS 21\_0      &       $9.043\pm0.015$	     
\\
AS 27\_1      &       $12.677\pm0.024$       
\\
AS 29\_1      &       $13.566\pm0.019$       
\\
AS 31\_1      &       $12.131\pm0.011$       
\\
P550\_C       &       $12.121\pm0.005$	     
\\
\hline
\end{tabular}
\end{table}
\onecolumn
\begin{table}
\caption{The logbook of the observations.}\label{logbooktab}
{\tiny
\begin{minipage}[top]{21truecm}
\begin{tabular}{ccccccccccccccc}
\hline
\hline
Galaxy & NGC & R.A. &  Dec & Type &$m_p$& Obs Date & Tel & Ncds & Ncmb & Exp T & Int. T & airm & 
Filt & seeing \\ 
\multicolumn{2}{c}{}&
\multicolumn{2}{c}{\hrulefill\ B1950.0 \hrulefill\ }&
&\rm mag&&&&&\rm sec&\rm sec&&&\rm arcsec\\
(1)  & (2)  & (3) & (4)  &  (5)   & (6)  & (7)  & (8)    & (9)   & (10)   & (11)  & (12) & (13)&(14)&(15) \\
\noalign{\smallskip}
\hline
\noalign{\smallskip}
VC0021&I3025&120749.80&+102800.0&	dS0 			& 14.75 &2000-03-27 & NTT &6&12&10&720&1.54&H&0.70\\
VC0033&I3032&120834.80&+143306.0&	dE nuc 			& 14.67 &2000-03-28 & NTT &10&12&6&720&1.38&H&0.75\\
VC0048&I3036&120942.10&+124559.0&	Sm			& 14.30 &2000-03-28 & NTT &10&18&6&1080&1.37&H&0.93\\
VC0067&I3044&121015.50&+141515.0&	dSc pec			& 13.98 &2000-03-28 & NTT &10&24&6&1440&1.41&H&0.73\\
VC0083&I3049&121100.60&+144530.0&	Im 			& 15.13 &2000-03-27 & NTT &6&18&10&1080&1.43&H&0.64\\
VC0162&I3074&121313.20&+105834.0&	Sd			& 14.41 &2000-03-28 & NTT &10&24&6&1440&1.71&H&0.74\\
VC0170&I3077&121323.80&+144240.0&	dS0			& 14.56 &2000-03-27 & NTT &10&12&6&720&1.73&H&0.71\\
VC0172&-&121327.60&+045542.0&		BCD			& 14.50 &2000-03-28 & NTT &6&17&10&1020&1.80&H&0.77\\
VC0216&I3097&121428.20&+094112.0&	dE		 pec	& 14.90 &2000-03-27 & NTT &10&15&6&900&1.86&H&0.90\\
VC0227&-&121441.40&+091312.0&		dE		 nuc	& 14.90 &2000-03-28 & NTT &10&18&6&1080&1.37&H&0.99\\
VC0275&I3118&121538.40&+094642.0&	dS0		  	& 14.54 &2000-03-28 & NTT &10&18&6&1080&2.13&H&0.87\\
VC0308&I3131&121618.00&+080818.0&	dS0		  nuc	& 14.30 &2000-03-27 & NTT &6&12&10&720&1.31&H&0.71\\
VC0437&-&121816.20&+174554.0&		dE		 nuc	& 14.54 &2000-03-28 & NTT &10&18&6&1080&1.46&H&0.80\\
VC0620&I3239&122037.20&+120011.0&	Sm		 	& 15.20 &2000-03-27 & NTT &10&18&6&1080&1.58&H&0.77\\
VC0688&N4353&122127.30&+080343.0&	 dSc	      		& 13.94 &2000-03-28 & NTT &6&12&10&720&1.42&H&0.87\\
VC0737&-&122206.60&+041636.0&		dS/BCD  		& 14.94 &2000-03-27 & NTT &6&18&10&1080&2.03&H&0.78\\
VC0750&-&122216.80&+070212.0&		dE		 nuc	& 14.95 &2000-03-27 & NTT &6&18&10&1080&1.49&H&0.66\\
VC0856&I3328&122324.60&+101948.0&	dE		 	& 14.25 &2000-03-27 & NTT &6&12&10&720&1.38&H&0.69\\
VC0951&I3358&122422.20&+115642.0&	dE/dS0  	     pec& 14.35 &2000-03-27 & NTT &10&6&6&360&1.52&H&0.80\\
VC0975&-&122438.30&+073223.0&		 Scd		 	& 13.58 &2000-03-28 & NTT &10&24&6&1440&1.67&H&0.90\\
VC1011&-&122456.50&+075514.0&		S0E		 	& 14.85 &2000-03-27 & NTT &10&18&6&1080&1.37&H&0.73\\
VC1036&N4436&122510.20&+123530.0&	dE/dS0  	     nuc& 13.68 &2000-03-27 & NTT &10&6&6&360&1.42&H&0.73\\
VC1047&N4440&122521.50&+123411.0&	 Sa		 bar	& 12.48 &2000-03-27 & NTT &10&2&6&120&1.42&H&0.73\\
VC1183&I3413&122651.00&+114230.0&	dS0		  nuc	& 14.37 &2000-03-27 & NTT &6&12&10&720&1.44&H&0.67\\
VC1392&I3459&122922.80&+122660.0&	dE/dS0  	     pec& 14.86 &2000-03-28 & NTT &10&18&6&1080&1.48&H&0.59\\
VC1491&I3486&123042.40&+130800.0&	dE		 nuc	& 15.24 &2000-03-27 & NTT &10&8&6&480&1.41&H&0.81\\
VC1499&I3492&123048.20&+130744.0&	dE		 pec	& 14.94 &2000-03-27 & NTT &10&8&6&480&1.41&H&0.81\\
VC1514&-&123104.80&+080818.0&		dE		 nuc	& 15.10 &2000-03-27 & NTT &6&18&10&1080&1.62&H&0.74\\
VC1528&I3501&123120.10&+133554.0&	dE		 	& 14.51 &2000-03-28 & NTT &10&6&6&360&1.44&H&0.74\\
VC1549&I3510&123143.20&+112051.0&	dE		 nuc	& 14.63 &2000-03-27 & NTT &6&12&10&720&1.31&H&0.64\\
VC1684&I3578&123407.80&+112242.0&	dS0		  	& 14.87 &2000-03-27 & NTT &6&12&10&720&1.34&H&0.78\\
VC1695&I3586&123422.80&+124800.0&	dS0		  	& 14.53 &2000-03-27 & NTT &6&12&10&720&1.34&H&0.61\\
VC1834&N4600&123749.40&+032338.0&	 S0		 nuc	& 13.47 &2000-03-28 & NTT &10&12&6&720&2.09&H&0.87\\
VC1895&-&123920.40&+094030.0&		dE		 	& 14.91 &2000-03-28 & NTT &10&12&6&720&2.10&H&0.87\\
VC1947&-&124023.30&+035701.0&		dE		 nuc	& 14.56 &2000-03-27 & NTT &6&12&10&720&2.04&H&0.93\\
VC2042&-&124407.20&+093448.0&		dE		 nuc	& 14.84 &2000-03-27 & NTT &6&18&10&1080&1.76&H&0.70\\
VC2050&I3779&124449.80&+122624.0&	dE		 nuc	& 15.20 &2000-03-27 & NTT &6&12&10&720&1.34&H&0.67\\
2MASX\tiny\footnote{\tiny 2MASX1J1211211+141438}&-&120848.10&+143120.0& Im/S&17.24&2000-03-28 & NTT &10&4&6&240&1.38&H&0.68\\
\noalign{\smallskip}
CCC045&-&124537.20&-405103.2&E	    &14.93&2000-03-27 & NTT &10&16&6&960&2.20&H&0.84\\
CCC059&-&124553.20&-405046.0&dE nuc &18.60&2000-03-27 & NTT &10&16&6&960&2.20&H&0.84\\
CCC094&-&124639.50&-410926.0&dS0 nuc&16.67&2000-03-27 & NTT &10&16&6&960&1.61&H&0.84\\
CCC095&-&124640.20&-411303.0&S0     &14.56&2000-03-27 & NTT &10&16&6&960&1.61&H&0.84\\
CCC096&-&124640.70&-411127.4&S0     &14.70&2000-03-27 & NTT &10&16&6&960&1.61&H&0.84\\
CCC104&-&124649.80&-410916.0&dE nuc &18.25&2000-03-27 & NTT &10&16&6&960&1.61&H&0.84\\
CCC113&-&124656.00&-405725.0&E	    &17.33&2000-03-26 & NTT &6&16&10&960&2.14&H&1.10\\
CCC119&-&124705.60&-405714.1&E	    &14.64&2000-03-26 & NTT &6&16&10&960&2.14&H&1.10\\
CCC122&N4706&124708.10&-410026.3&S0a&14.18&2000-03-26 & NTT &6&16&10&960&2.14&H&1.10\\
CCC125&-&124710.40&-405917.0&dE nuc &17.14&2000-03-26 & NTT &6&16&10&960&2.14&H&1.10\\
CCC136&-&124725.70&-410138.4&E      &16.25&2000-03-28 & NTT &10&8&6&480&1.71&H&0.84\\
CCC142&-&124731.30&-410156.0&Sm     &18.40&2000-03-28 & NTT &10&8&6&480&1.71&H&0.84\\
CCC150&-&124738.00&-410130.0&dE nuc &18.23&2000-03-28 & NTT &10&8&6&480&1.71&H&0.84\\
CCC153&-&124741.60&-410210.0&dE nuc &18.24&2000-03-28 & NTT &10&8&6&480&1.71&H&0.84\\
CCC157&-&124747.30&-410351.0&Sm     &18.23&2000-03-28 & NTT &10&8&6&480&1.71&H&0.84\\
CCC205&-&124901.60&-404321.0&S0     &15.95&2000-03-28 & NTT &6&6&10&360&1.90&H&0.78\\
CCC216&-&124915.80&-410454.0&dE nuc &18.22&2000-03-27 & NTT &10&16&6&960&1.70&H&0.78\\
CCC222&-&124926.20&-410402.9&dSc    &14.74&2000-03-27 & NTT &10&16&6&960&1.70&H&0.78\\
CCC226&N4743&124929.30&-410708.6&S0a&14.08&2000-03-27 & NTT &10&16&6&960&1.70&H&0.78\\
CEG050\tiny\footnote{\tiny [BCS89] 050}&-&124828.10&-410903.0&dE&? &2000-03-27 & NTT &10&16&6&960&2.20&H&0.84\\
\noalign{\smallskip}			  
VC0010&I3017&120651.70&+135110.0&BCD       &14.75&1999-12-19 & TNG &6&8&10&480&1.78&H&1.50\\
VC0608&N4322&122029.70&+161058.0&dE     nuc&14.94&1999-12-20 & TNG &6&8&10&480&1.90&H&1.30\\
VC0745&N4366&122214.40&+073748.0&dE     nuc&14.67&1999-12-20 & TNG &4&5&10&200&1.22&H&1.09\\
VC0786&I3305&122243.80&+120754.0&dE     nuc&15.11&1999-12-20 & TNG &6&9&10&540&1.25&H&1.21\\
VC0965&I3363&122431.20&+125006.0&dE     nuc&15.40&1999-12-20 & TNG &6&8&10&480&1.77&H&1.27\\
VC1073&I794&122536.50&+122211.0&	dE     nuc&14.23&1999-12-20 & TNG &6&8&10&480&1.08&H&0.90\\
VC1078&-&122539.00&+100224.0&	dE     ?  &15.30&1999-12-19 & TNG &6&12&10&720&1.20&H&1.06\\
VC1122&I3393&122609.60&+131130.0&dE     nuc&14.60&1999-12-19 & TNG &4&8&10&320&1.76&H&1.90\\
VC1173&-&122643.00&+131516.0&	dE     nuc&16.06&1999-12-20 & TNG &6&8&10&480&1.58&H&1.76\\
VC1254&-&122732.80&+082103.0&	dE     nuc&15.51&1999-12-20 & TNG &6&8&10&480&1.15&H&0.88\\
VC1308&I3437&122814.40&+113700.0&dE     nuc&15.64&1999-12-20 & TNG &4&8&10&320&1.46&H&1.17\\
VC1348&I3443&122843.90&+123628.0&dE     pec&15.87&1999-12-20 & TNG &4&8&10&320&1.36&H&1.12\\
VC1386&I3457&122919.20&+125600.0&dE     nuc&14.32&1999-12-19 & TNG &4&10&10&400&1.51&H&1.13\\
VC1453&I3478&123012.80&+142819.0&dE     nuc&14.34&1999-12-19 & TNG &4&8&10&320&1.63&H&1.48\\
VC1491&I3486&123042.40&+130800.0&dE     nuc&15.24&1999-12-19 & TNG &6&8&10&480&1.36&H&1.10\\
\noalign{\smallskip}
97073&-&114020.75&+201438.1&pec&15.60&1999-12-19 & TNG &6&8&10&480&1.67&H&1.27\\
97087&-&114113.19&+201449.1&pec&14.30&1999-12-20 & TNG &3&11&10&330&1.62&H&1.31\\
\noalign{\smallskip}
1ZW018&-&093030.10&+552747.0&BCD&16.08&1999-12-20 & TNG &6&8&10&480&1.28&H&1.10\\
\noalign{\smallskip}
\end{tabular}
\end{minipage}
}
\end{table}
\twocolumn
\normalsize

\subsection{Image reduction procedures}
The multiplicative system response, or flat-field (FF), was derived for
the ESO-NTT observations from a set of dome exposures which allow to
remove both the dependence of the dark current level from the
illumination of the array and the additive contributions.  For the TNG 
observations, since dome exposures could not be taken, the FF was obtained
averaging, and normalizing to their median counts, a large number 
($\gtrsim30$) of sky frames taken throughout the night, with mean levels
differing by less than 5\%.\\
The image reduction procedure was as follows.  For each target frame the
sky contribution was determined and subtracted.  This was done by
combining, with a median sigma clipping algorithm, as many as possible
contiguous sky exposures, unless their count level differed by more than
10\% from the target frame. In the case of mosaics with the source
always in the field, all frames were treated as sky frames.  The median
sigma clipping algorithm is necessary to remove unwanted star and galaxy
images in the resulting sky frames.  The sky frame was first normalized
to its median, then multiplied by the median counts of the individual
target frames.  Finally, the rescaled frame was subtracted from the
target observation.  Such a procedure accounts for temporal variations
in the sky level, but introduces an additive offset which is
subsequently removed (see below).  The sky-subtracted target frames were
then divided by the FF frame.  Each of the corrected frames was analyzed
for low-spatial-frequency gradients, and if necessary, fitted with a
two-dimensional 3 degree polynomial which was then subtracted.  If this
process was not effective in removing the spatial gradients, the
corresponding frames were rejected from further analysis.  The corrected
frames were then aligned using field stars and combined with a median
filter with sigma clipping, which allows bad pixel removal.\\
Finally the residual sky background in the combined frame was determined as
the mean number of counts measured in regions of ``empty'' sky, and it was
subtracted from the frame.\\
All image reduction and analysis was performed in the IRAF environment
and relied on the STSDAS package\footnote{IRAF is the Image Analysis and
Reduction Facility made available to the astronomical community by the
National Optical Astronomy Observatories, which are operated by AURA,
Inc., under contract with the U.S. National Science Foundation. STSDAS
is distributed by the Space Telescope Science Institute, which is
operated by the Association of Universities for Research in Astronomy
(AURA), Inc., under NASA contract NAS 5--26555.}, and on GALPHOT
(developed for IRAF--STSDAS mainly by W. Freudling, J. Salzer, and M.P.
Haynes and adapted by us to handle NIR data).\\ 
The final images, with superposed isophotes, are shown in Figs. 7-8.\\
\section{Profile decomposition procedures}
The 2-dimensional light distribution of each galaxy was fitted with
elliptical isophotes, using a procedure based on the task ${\it
ellipse}$, (STSDAS ${\it ISOPHOTE}$ package; Jedrzejewski, 1987, and
Busko, 1996), which allows the interactive masking of unwanted
superposed sources.  Starting from an interactively centered ellipse,
the fit maintains as free parameters the ellipse center, ellipticity,
and position angle. The ellipse semi-major axis is incremented by a
fixed fraction of its value at each step of the fitting procedure.  The
routine halts when the surface brightness found in a given corona equals
the sky rms, and then restarts decrementing the initial semi-major axis
toward the center.  Isophotes whose rms is greater than their mean value
are discarded. The fit fails to converge for some galaxies with very
irregular light distributions.  In these cases we kept fixed one or more
of the ellipse parameters.\\ 
The resulting radial light profiles were fitted using one of four models
of light distributions:\\ 
1) a de Vaucouleurs $r^{1/4}$ law (de Vaucouleurs, 1948);\\
2) an exponential law;\\
3) a ``mixed'' profile consisting of the sum (in flux) of an exponential
law, dominating at large radii (``disk''), and an exponential or a de 
Vaucouleurs $r^{1/4}$ law, dominating at small radii (``bulge'');\\
4) a ``truncated'' profile consisting of an exponential or a de 
Vaucouleurs $r^{1/4}$ law, truncated by a steeper exponential law beyond
a certain critical radius $r_{\rm t}$, according to either of the following:
\begin{eqnarray}
\lefteqn{I(r)=c_1\cdot {exp\big[-\frac{1}{c_2}({r-r_{\rm t}}-|{r-r_{\rm t}}|)\big]}\cdot{}}
\nonumber\\
& &{}\cdot{exp\big[-\frac{1}{c_3}({r-r_{\rm t}}+|{r-r_{\rm t}}|\big]}~{\rm(Truncated~exponential)}\nonumber\\
\lefteqn{I(r)=c_1\cdot{exp\big[-\frac{1}{c_2}({r^{1/4}-r_{\rm t}^{1/4}}-|{r^{1/4}-r_{\rm t}^{1/4}}|)\big]}\cdot{}}
\nonumber\\
&&{}\cdot{exp\big[-\frac{1}{c_3}({r-r_{\rm t}}+|{r-r_{\rm t}}|\big]}~{\rm(Tr.~DeVaucouleurs)}\nonumber
\end{eqnarray}
For pure de Vaucouleurs and exponential laws, the fit was performed using
a weighted least squares method. For the mixed and truncated profiles, 
the fit was performed using the Levemberg-Marquardt algorithm implemented
in the task ${\it nfit1d}$ (STSDAS ${\it FITTING}$ package).  
This algorithm is implemented within an interactive procedure
which requires some initial set of parameters i.e. 4 markers delimiting
the outer or exponential dominated region, and the inner or bulge
dominated region. The former is fitted with an exponential law.  For
mixed profiles, the external exponential fit is extrapolated to the
inner region and subtracted. The resulting inner profile 
is then fitted either with an exponential or a de Vaucouleurs $r^{1/4}$ 
law, according to a $\chi^2$ test. Fitting parameters are then assumed 
as initial guess for the Levemberg-Marquardt algorithm.
For truncated profiles, the inner region is fitted either with
an exponential or a de Vaucouleurs $r^{1/4}$ law, according to a
$\chi^2$ test, and the fitting parameters are then used as initial guess,
along with the external exponential slope and the inner edge of the 
outer region as $r_{\rm t}$.\\
The fits are performed from a radius equal to twice the seeing disk, out
to the outermost significant isophotes.\\
Total magnitudes $H_{\rm T}$ are then obtained by adding to the flux measured
within the outermost significant isophote the flux extrapolated to
infinity along the fitted profile. The $1-\sigma$ error attached 
to the total magnitude {$H_{\rm T}$} combines the statistical error on 
the flux at the outermost isophote with that on the fit parameters.\\
The effective radius $r_{\rm e}$ (the radius containing half of the total
light) and the effective surface brightness $\mu_{\rm e}$ (the mean surface
brightness within $r_{\rm e}$) of each galaxy are ``empirically'' computed
(see Paper V).
The relative errors are obtained combining the uncertainty on 
$H_T$, as described above, with the scatter $\sigma_r$ along the 
integrated-light growth curve.\\
Finally we compute other useful parameters: the concentration index
($C_{31}$), defined in de Vaucouleurs (1977) as the model--independent
ratio between the radii that enclose 75\% and 25\% of the total light
$H_T$, and, for galaxies fitted with a two component model, the bulge to
total flux ratio ($B/T$).\\ 
The derived surface brightness profiles are shown in Fig. 9:
each galaxy is labelled with a prefix denoting the telescope (N00 for
ESO-NTT or G99 for TNG), followed by its catalogue name and by the type
of decomposition (see Tab.\ref{tabdecomp}).  
\section{Results}
The results of the present work are summarized in Tab.\ref{tabdecomp},
as follows:\\ 
Column  1: VCC (Binggeli et al. 1985) or CCC (Jerjen \& Dressler 1997) or
 CGCG (Zwicky et al. 1961-68) designation.\\
Column 2: adopted filter (B or H). \newline 
Column 3: type of decomposition: D = pure de Vaucouleurs; E = pure
exponential; M = mixed; T = truncated. \newline 
Column 4: nucleus: Y=present, not fitted; N=absent; B=extended, fitted as 
a bulge\newline
Column 5: type of decomposition of the bulge: D = de Vaucouleurs; E =
exponential. \newline 
Column 6: effective radius of the fitted bulge component ($r_{\rm ebf}$)
in arcsec.\newline 
Column 7: effective surface brightness of the fitted bulge component
($\mu_{\rm ebf}$) in \marc.\newline 
Column 8: effective radius of the fitted disk component ($r_{\rm edf}$)
in arcsec.\newline 
Column 9: effective surface brightness of the fitted  disk component
($\mu_{\rm edf}$) in \marc.\newline 
Column 10: effective radius of the fitted outer exponential component ($r_{\rm eout}$)
in arcsec given for truncated profiles.\newline 
Column 11: effective surface brightness of the fitted outer exponential
component ($\mu_{\rm eout}$) in \marc.\newline 
Column 12-13: total effective radius ($r_{\rm e}$) and associated uncertainty in
arcsec.\newline  
Column 14-15: total effective surface brightness ($\mu_{\rm e}$) and associated
uncertainty in \marc.\newline  
Column 16-17: total magnitude ($m_{\rm T}$) extrapolated to infinity and associated
uncertainty.\newline  
Column 18-19: concentration index ($C_{31}$) and associated 
uncertainty.\newline
Column 20: bulge to total flux ratio ($B/T$).\newline 
\newline 

\section{Analysis}
The analysis presented in this Sect. is based on 818
galaxies listed in Tab.\ref{completeness}. Assuming  an average $B-H=3$ mag,
the optical completeness levels given in Sect. 2 translate
into 100\% completeness at $Log~L_{\rm H\odot}>10$ for the Coma supercluster,
100\% at $Log~L_{\rm H\odot}>9.4$ and
50\% at $Log~L_{\rm H\odot}>8.6$ for the Virgo cluster.
The 50\% unobserved Virgo galaxies with $16.0>m_{\rm p}>14.0$ mag includes 
objects whose surface brightness, as judged on the DSS plates, was
fainter than what we could expect to detect with a 4m telescope
in one-hour integration.\\

\subsection{The frequency of profile decompositions}
\begin{figure}
\centerline{
\includegraphics[height=10truecm,width=10truecm]{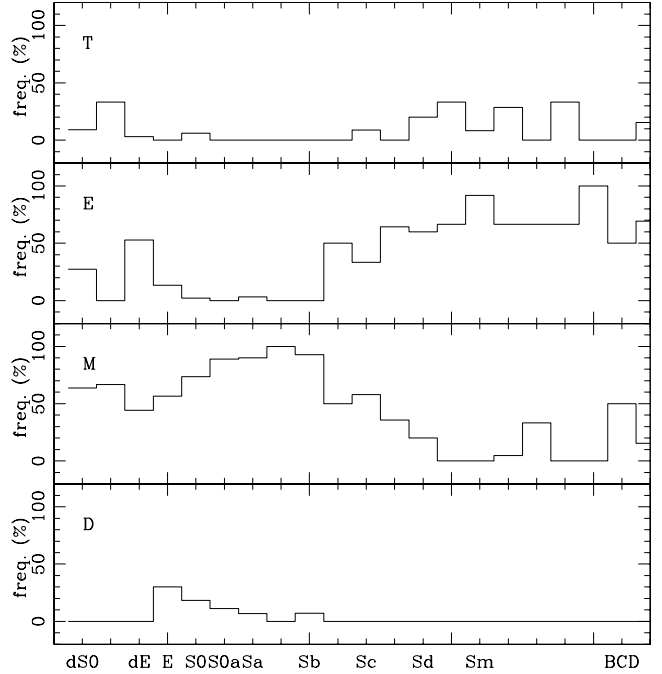}}
\caption{The fraction of pure de Vaucouleurs, mixed, pure
exponential and truncated profiles along the Hubble
sequence for galaxies in the Virgo cluster.}\label{freq_hist}
\end{figure}
\begin{figure}[!h]
\centerline{
\includegraphics[height=8truecm,width=8truecm]{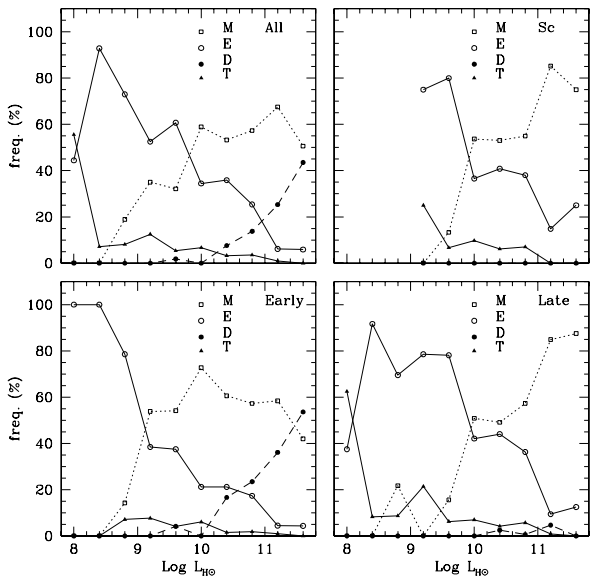}}
\caption{The fraction of pure de Vaucouleurs, mixed, exponential and 
truncated profiles as a function of the NIR luminosity among all galaxies
(top-left), among the Sbc-Sd (top-right), among early types (dE-E-S0a)
(bottom-left) and among spirals (Sa-BCD) (bottom-right).
The completeness levels of 100 \% and 50 \% for the Virgo cluster, 
computed assuming an average $B-H=3$ mag, are at 9.4 and 8.6 
$Log~L_{\rm H\odot}$ respectively.
}\label{freq_hlum}
\end{figure}
We consider the distribution of profile decompositions along the
Hubble sequence only for galaxies in the Virgo cluster, for which the
morphological classification is most reliable. This is shown in Fig.
\ref{freq_hist}.  It is apparent that pure de Vaucouleurs profiles are
present only in 40\% of Es and in 30\% of S0s. Their contribution drops
to zero both for later types and for the early type dwarfs.  The exponential
profiles are nearly absent among early type giant systems up to Sab, while
their frequency is high (44\%) in dwarf E+S0s, and increases from 40\%
(Sc) to almost 100\% for later types.  Mixed (M) decompositions dominate
among dwarf E and S0s (50\%) and giant Es (50\%), increasing up to 90\% among
Sb galaxies, then drop to zero for later types.  Truncated profiles (T) are
rare (their frequency is always $< 35\%$), and are absent from giant
early-type galaxies up to Sc spirals.\\
Fig. \ref{freq_hlum} shows the relative fraction of profile
decompositions plotted as a function of the H band luminosity ($Log~
L_{\rm H}/$\lsol$=11.36-0.4H_{\rm T}+2Log~D$ (D in Mpc)) for the 818 objects in the
Virgo+Coma sample (top-left panel), for the early type
(dE-E-S0a)(bottom-left panel), late type (Sa-BCD)(bottom-right panel)
and Sc-Sd galaxies alone (top-right panel). All panels show similar trends,
indicating that the dependence of the frequency of profile
decompositions on luminosity is independent of the morphological type.
The fraction of pure de Vaucouleurs profiles strongly increases with the
H luminosity, being absent for $L_{\rm H}<10^{9.5}$ \lsol, a luminosity range
where the pure exponential profiles dominate, since their frequency
clearly anti-correlates with luminosity. At the faintest luminosities,
however truncated profiles are abundant among late-type galaxies.  The
frequency of mixed profiles increases monotonically with luminosity
among late-type galaxies, while it reaches a maximum at
$L_{\rm H}\sim~10^{10}$ \lsol for the early-type ones, because for higher
luminosities these galaxies have increasingly more frequently pure 
de Vaucuoleurs profiles.  

\subsection{The light concentration parameter $C_{31}$}
\begin{figure}[!h]
\centerline{
\includegraphics[height=8truecm,width=8truecm]{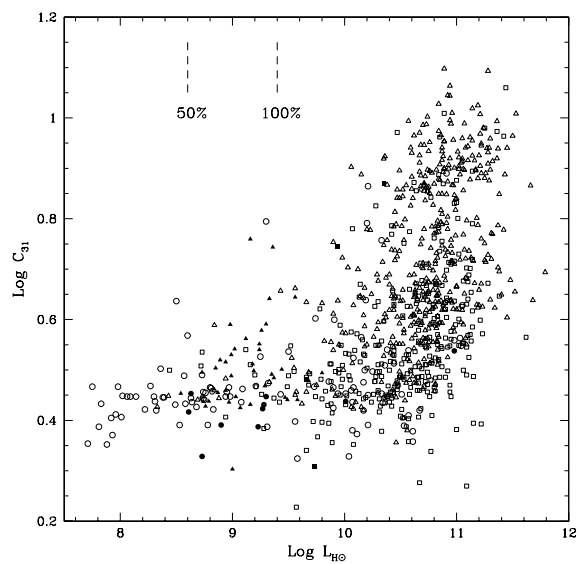}}
\caption{The dependence of $C_{31}$ on luminosity. Points are coded
in three classes of Hubble type (dE-S0a=triangles; Sa-Scd=squares;
Sd-BCD=circles). Measurements reported in this paper are given with 
filled symbols. Open symbols are from Paper V. Two completeness levels 
for the Virgo cluster, computed assuming an average $B-H=3$ mag, are 
indicated with dashed lines. }\label{c31_hlum}
\end{figure}

Fig. \ref{c31_hlum} shows the remarkable dependence of $C_{31}$ on
luminosity found by Scodeggio et al. (in preparation) and extended here
to comprise dwarf galaxies.  We confirm that high $C_{31}$ 
(cusps+extended haloes) are almost completely absent at 
$L_{\rm H}<10^{9.5}$\lsol.  Faint galaxies cluster around $C_{31}=2.80$, 
which is the expected value for pure exponential profiles. This result is 
completely independent from galaxy morphology.  On the other end
high $C_{31}$ (bulge-dominated) objects are present only at
high luminosity $L_{\rm H}>10^{10.5}$\lsol. These are a mixture of giant E
and Early-type spirals.  There exist however a significant class of
high-luminosity (giant), low $C_{31}$ (bulge-less) galaxies which
appears to be confined to Sc galaxies. 

\subsection{Color gradients}
\begin{figure}[!b]
\centerline{
\includegraphics[height=8truecm,width=8truecm]{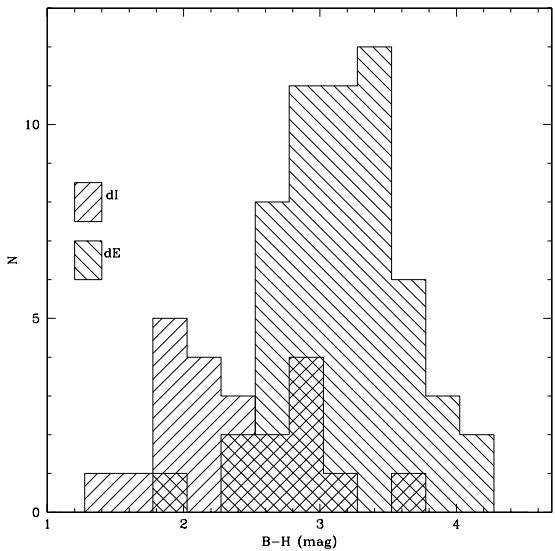}}
\caption{The distribution of dE/dS0 and dI 
in bins of B-H.}\label{histcol}
\end{figure}

Only 22 dE/dS0 in Virgo analyzed in this paper have either B-V or B-H
color profile (see Fig. 10). Nine among these 22 (41\%) have no
radial color gradients. Another 9 have a red central excess 
consistent with an age or metallicity gradient toward the center.  
The remaining 4 (VCC 781, 951, 1499, 1684, representing a 
non-negligeable 18\% of our sample) have instead a blue central excess 
consistent with a nuclear post star-burst phase (see below the discussion 
on VCC 1499). In spite of the paucity of the available data,
we find that the color of the central excess correlates with the global color
($<B-V>_{\rm B excess}=0.61$, $<B-V>_{\rm R excess}=0.82$) and that the 4 $B_{\rm excess}$
objects have on average slightly lower $<L_{\rm H}>=8.9$ than those with
$R_{\rm excess}$ ($<L_{\rm H}>=9.1$).\\
This evidence is consistent with the analysis by Kormendy \& Djorgovski 
(1989) (see their Fig. 4).\\
We also find that the color distribution of dEs overlaps with that 
of dIs (see Fig.\ref{histcol}), though the mean colors of the two 
groups differ significantly.\\
Excluding galaxies with $m_{\rm p}>16.0$, because we don't have measurements 
of dEs fainter than this limit, we explored the possible continuity in 
the structural parameters of dEs and dIs, as proposed by Sung et al. (1998).
dEs (9 objects) and nucleated dE-Ns (27 objects) are indistinguishable 
from each other both in colors ($<B-H>\sim3.1$) and in $C_{31}$ 
($<C_{31}>\sim3.4$). The only subclass of dEs with significantly 
bluer colors and lower $C_{31}$ are dE-pecs (6 objects) which 
have $<B-H>\sim2.7$ and $<C_{31}>\sim2.7$, consistent with 
$<B-H>\sim2.5$ and $<C_\mathit{31}>\sim2.5$ of dIs (16 objects).\\
\begin{figure}[!t]
\centerline{
\includegraphics[height=7truecm,width=7truecm]{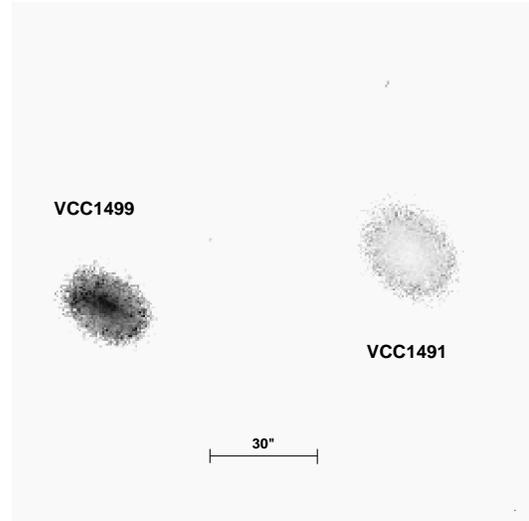}}
\caption{B-H color map of VCC1491-1499 (white = red, black = blue).
Grey levels span from B-H=2.35 (darkest) to B-H=3.75. North is up, East to the left.}\label{gradient}
\end{figure}
\begin{figure}[!t]
\centerline{
\includegraphics[height=8truecm,width=8truecm]{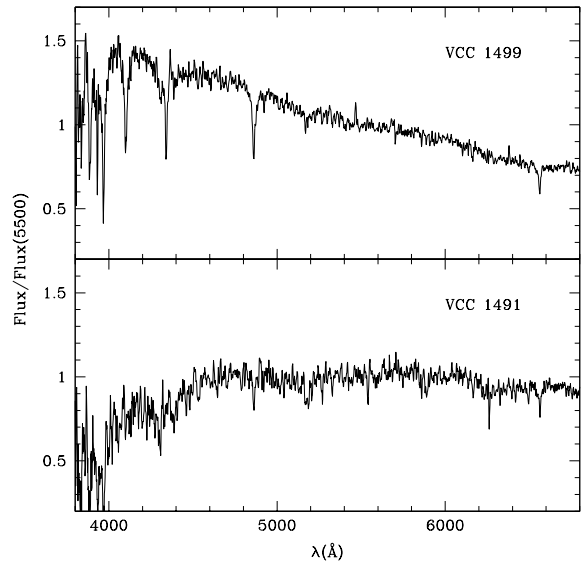}}
\caption{Long-slit spectra of VCC1491-1499.
The flux is normalized at $5500 \rm\AA$.}\label{spectra}
\end{figure}
An illustrative and meaningful example of the wide range spanned in color 
and color gradient by dEs is offered by the two galaxies VCC1491 and 1499 
which happen to lie 1.5 arcmin apart in the same frame (see Fig. 7).
The two have V mag differing by 0.01 mag, thus they are indistinguishable 
galaxies in all respects. However VCC1491 is almost as red (B-V=0.82, 
B-H=3.42) as a giant elliptical, while on the opposite VCC1499 (dE-pec) 
is almost as blue (B-V=0.49, B-H=2.67) as a typical dI. Moreover VCC1499 
shows a strong blue central excess, while VCC1491 presents a shallower red
gradient toward its centre (see Fig. \ref{gradient} and profiles in 
Fig. 10).
Using the Carelec spectrograph (Lemaitre et al. 1990) attached to the 
OHP 1.93m telescope we obtained in february 2000 long slit spectra for 
the two objects, shown in Fig. \ref{spectra}.
The spectral signatures of the two galaxies are significantly different:
1491 resembles a typical dE galaxy, while 1499 has a much bluer continuum 
and strong Balmer absorption lines (E.W. $H_\delta \sim 8 \rm \AA$) 
typical of E+A galaxies which have experienced an intense burst of star 
formation ended about 1-2 Gyrs ago (Poggianti \& Barbaro, 1996).\\

\section{Summary and Conclusions}
We obtained near-infrared H-band profile decompositions
for 75 galaxies taken primarily among dwarf galaxies in the Virgo cluster.
Adding these new observations to the ones similarly taken in the Virgo,
Coma and A1367 clusters and in the ``Great Wall'' (Paper V), we gathered 
H-band data for 818 galaxies.
These include all galaxies brighter than m$\rm _p =15.7$ in the Coma region,
corresponding to $M_{\rm p}<-19.2$ ($\mu$=34.9) and 94\% of galaxies brighter than 
m$\rm _p =14.0$ in the Virgo cluster, corresponding to $M_{\rm p}<-17.2$ 
($\mu$=31.2), thus the observations of giant galaxies are complete.
Considering only the Virgo cluster, we also covered 30\% of galaxies in 
the interval $14<m\rm_p<16.0$ corresponding to $-17.2<M_{\rm p}<-15.2$, thus to 
the transition region between giant and dwarf galaxies (see Sandage, 
Binggeli \& Tammann 1985). The completenes in the same magnitude range 
increases from 30 to 47\% if one considers the ISO sample only.
The studied sample is representative of all Hubble types, including dE and Im,
and spans 4 orders of magnitude in luminosity.\\ 
We model the surface brightness profiles of the studied galaxies with 
either a de Vaucouleurs $r^{1/4}$ law (D), an exponential law (E), a 
combination of the two (M), or with a profile that is truncated at the 
periphery (T). Using the fitted quantities we find that: \\ 
1) Less than 50\% of the giant elliptical galaxies have pure D profiles.\\ 
The majority of giant galaxies (E to Sb) is best represented by a M profile.
Scd-BCD galaxies have pure exponential profiles. \\ 
2) Most dwarf galaxies (independently from their detailed morphological type)
follow exponential profiles or truncated decompositions.\\
3) The type of decomposition is a strong function of the total H band
luminosity ($10^8<L_{\rm H}<10^{11.5}$ \lsol), irrespective of the galaxy Hubble
classification: the fraction of pure exponential profiles decreases with 
increasing luminosity, while that of M ones increases with luminosity.  
Truncated profiles are characteristic of the lowest luminosity galaxies. 
Pure D profiles are absent at low luminosities $L_{\rm H}<10^{10}$ \lsol and 
become dominant above $10^{11}$ \lsol.\\
4) The light concentration index $C_{31}$ (presence of central 
cusps and extended outer haloes) is a strong non-linear function of the 
total luminosity, irrespective of the Hubble classification: 
dwarf systems have low $C_{31}$, typical of exponential disks;
high $C_{31}$, characteristic of conspicuous bulges,  are found only at
the highest luminosities. There exist however a class of bulge-less, high
luminosity galaxies. These are giant Sc's.\\
5) dE galaxies have mildly redder colors and higher $C_{31}$ than dIs.
The only subclass of early-type dwarfs having structural parameters 
indistinguishable from those of late-type dwarfs seems to be that of dE-pec,
which therefore represents the possible missing link between dEs and dIs.
This is supported by the evidence of post-starburst activity found in the 
dE-pec VCC1499.\\
The results summarized in points 1) through 4) should not suffer 
from selection biases since at low-luminosities, where our sample is
severely incomplete, we observed primarily the highest surface brightness 
galaxies. Thus a bias, if any is present, should select in favour of high 
$C_{31}$ galaxies with D or M profiles, because at comparable 
luminosities these objects have higher central surface brightness than 
those with low $C_{31}$ and E or T profiles.\\
Summarizing, points 1-4 indicate that the frequency of occurrence of
relevant cusps and extendend luminous haloes, absent among low-mass
galaxies, increases significantly with increasing mass.  This is
consistent with the monolithic collapse scenario (Sandage 1986)
provided that the collapse efficiency scales with mass (Gavazzi \&
Scodeggio 1996).  If, otherwise, merging is invoked as the mechanism
for building galaxies of increasing mass, a problem arises: while
extended haloes are naturally produced as remnants of mergers between
stellar disks, central high-brightness cusps require that the mergers
occur in the presence of a gaseous phase (Hernquist et al. 1993). If
this were the case, however, cusps (bulges) would be composed of
younger stellar populations than it is generally observed.

\acknowledgements{We thank C. Bonfanti for the reduction and analysis of
OHP spectra of two galaxies}

\onecolumn
\begin{landscape}
\tiny
\begin{longtable}{cccccccccccccccccccc}
\caption{Photometric parameters of the target galaxies.}\label{tabdecomp}\\
\hline
\hline
\multicolumn{5}{c}{}&
\multicolumn{2}{c}{\hrulefill\ Bulge \hrulefill\ }&
\multicolumn{2}{c}{\hrulefill\ Disk exp \hrulefill\ }&
\multicolumn{2}{c}{\hrulefill\ Outer exp \hrulefill\ }\\
Galaxy &  Filter &  dec & Nuc & Bdec & $r_{\rm ebf}$ & $\mu_{\rm ebf}$ & $r_{\rm edf}$ & $\mu_{\rm edf}$ & 
$r_{\rm eout}$ & $\mu_{\rm eout}$ & $r_{\rm e}$  & $\pm$ & $\mu_{\rm e}$ & $\pm$ & $M_{\rm T}$ & $\pm$  & $C_{31}$ & $\pm$&  B/T \\  
&   &    &      &      & arcsec & mag/$\Box$"   & arcsec & mag/$\Box$"  & arcsec & mag/$\Box$"  &
\multicolumn{2}{c}{arcsec} &
\multicolumn{2}{c}{ mag/$\Box$" } &
\multicolumn{2}{c}{ mag } & & &\\    
(1)  & (2)  & (3) & (4)  &  (5)   & (6)  & (7)  & (8)    & (9)   & (10)   & (11)  & (12)   & (13)  & (14)&  (15)
 & (16) & (17)   & (18)   &  (19) &(20)  \\
\\
\hline
\\
\endfirsthead
\caption{continued}\\
\hline
\multicolumn{5}{c}{}&
\multicolumn{2}{c}{\hrulefill\ Bulge \hrulefill\ }&
\multicolumn{2}{c}{\hrulefill\ Disk exp \hrulefill\ }&
\multicolumn{2}{c}{\hrulefill\ Outer exp \hrulefill\ }\\
Galaxy &  Filter &  dec & Nuc & Bdec & $r_{\rm ebf}$ & $\mu_{\rm ebf}$ & $r_{\rm edf}$ & $\mu_{\rm edf}$ & 
$r_{\rm eout}$ & $\mu_{\rm eout}$ & $r_{\rm e}$  & $\pm$ & $\mu_{\rm e}$ & $\pm$ & $M_{\rm T}$ & $\pm$  & $C_{31}$ & $\pm$&  B/T \\  
&   &    &      &      & arcsec & mag/$\Box$"   & arcsec & mag/$\Box$"  & arcsec & mag/$\Box$"  &
\multicolumn{2}{c}{arcsec} &
\multicolumn{2}{c}{ mag/$\Box$" } &
\multicolumn{2}{c}{ mag } & & &\\    
(1)  & (2)  & (3) & (4)  &  (5)   & (6)  & (7)  & (8)    & (9)   & (10)   & (11)  & (12)   & (13)  & (14)&  (15)
 & (16) & (17)   & (18)   &  (19) &(20)  \\
\\
\hline
\\
\endhead
\hline
\endfoot
VC0010&H&T&N&E&-&-&8.63&18.05&5.74&17.23&7.0&1.0&17.86&0.28&12.76&0.07&2.65&0.03&0.59      	 
\\
\noalign{\smallskip}
VC0021&H&M&N&E&4.50&19.49&13.67&19.57&-&-&12.0&1.2&19.16&0.18&12.33&0.07&3.31&0.01&0.09    	 
\\
\noalign{\smallskip}
VC0033&H&E&N&-&-&-&9.17&18.90&-&-&8.7&1.1&18.77&0.23&12.24&0.07&2.79&0.02&0.00             	 
\\
\noalign{\smallskip}
VC0048&H&E&N&-&-&-&18.55&20.58&-&-&19.2&1.1&20.65&0.10&12.72&0.06&2.70&0.03&0.00           	 
\\
\noalign{\smallskip}
VC0067&H&M&N&E&11.48&19.97&66.67&21.60&-&-&53.8&6.0&20.98&0.19&11.27&0.08&5.55&0.14&0.18   	 
\\
\noalign{\smallskip}
VC0083&H&E&N&E&-&-&15.47&20.56&-&-&15.4&1.1&20.55&0.13&12.98&0.06&2.84&0.01&0.00           	 
\\
\noalign{\smallskip}
VC0162&H&E&N&-&-&-&50.45&20.19&-&-&62.2&2.0&20.91&0.06&11.48&0.06&2.44&0.02&0.00           	 
\\
\noalign{\smallskip}
VC0170&H&M&N&E&4.37&20.75&17.41&19.77&-&-&17.0&1.1&19.71&0.11&11.98&0.06&2.80&0.02&0.02    	 
\\
\noalign{\smallskip}
VC0172&H&E&N&-&-&-&13.73&19.53&-&-&13.5&1.1&19.60&0.15&12.68&0.06&2.80&0.01&0.00           	 
\\
\noalign{\smallskip}
VC0216&H&T&Y&E&-&-&16.86&19.53&12.65&19.24&12.8&1.2&19.30&0.17&12.21&0.06&2.76&0.00&0.28   	 
\\
\noalign{\smallskip}
VC0227&H&E&Y&-&-&-&20.33&20.38&-&-&21.0&1.3&20.51&0.11&12.33&0.06&3.17&0.02&0.00           	 
\\
\noalign{\smallskip}
VC0275&H&E&N&-&-&-&19.80&20.35&-&-&18.9&1.1&20.19&0.10&12.54&0.06&3.10&0.01&0.00           	 
\\
\noalign{\smallskip}
VC0308&H&T&B&D&143.68&21.77&-&-&15.77&18.81&17.1&1.2&19.10&0.12&11.00&0.06&3.01&0.01&0.54  	 
\\
\noalign{\smallskip}
VC0437&H&M&Y&E&7.69&18.94&28.63&20.06&-&-&20.0&1.5&19.12&0.13&11.22&0.06&4.38&0.11&0.21    	 
\\
\noalign{\smallskip}
VC0608&B&T&Y&E&-&-&16.47&22.73&12.49&22.13&15.7&1.2&22.83&0.12&15.29&0.07&2.67&0.02&0.59   	 
\\
VC0608&H&E&Y&-&-&-&17.26&19.34&-&-&17.7&1.1&19.43&0.11&11.87&0.06&2.66&0.01&0.00           	 
\\
\noalign{\smallskip}
VC0620&H&E&N&-&-&-&17.02&20.29&-&-&21.8&1.2&21.16&0.09&13.02&0.07&2.61&0.02&0.00           	 
\\
\noalign{\smallskip}
VC0688&H&T&N&D&15021.10&23.97&-&-&9.94&16.55&18.0&1.1&18.65&0.09&11.00&0.07&2.03&0.01&0.58 	 
\\
\noalign{\smallskip}
VC0737&H&T&N&E&-&-&25.52&19.58&11.32&18.56&14.6&1.1&19.33&0.12&12.72&0.06&2.13&0.04&0.34   	 
\\
\noalign{\smallskip}
VC0745&B&M&N&E&6.53&22.02&24.36&23.69&-&-&16.3&1.4&22.47&0.15&14.95&0.06&4.12&0.09&0.29    	 
\\
VC0745&H&E&?&-&8.99&18.26&-&-&-&-&8.8&1.1&18.31&0.23&11.99&0.06&2.94&0.05&0.00             	 
\\
\noalign{\smallskip}
VC0750&H&M&Y&E&5.25&19.47&22.58&20.49&-&-&18.2&1.5&19.84&0.15&11.95&0.06&3.90&0.02&0.15    	 
\\
\noalign{\smallskip}
VC0781&B&M&Y&E&4.52&21.94&15.52&22.43&-&-&13.2&1.3&21.93&0.17&14.98&0.07&4.08&0.07&0.17
\\
\noalign{\smallskip}
VC0786&B&M&Y&E&10.25&23.32&32.13&23.68&-&-&27.8&3.0&23.20&0.15&14.84&0.11&3.84&0.24&0.18   	 
\\
VC0786&H&E&Y&-&-&-&16.20&19.58&-&-&15.6&1.4&19.66&0.17&12.21&0.06&3.28&0.01&0.00           	 
\\
\noalign{\smallskip}
VC0856&H&E&Y&-&-&-&14.83&18.98&-&-&14.3&1.2&18.93&0.15&11.27&0.06&2.76&0.04&0.00           	 
\\
\noalign{\smallskip}
VC0916&B&M&N&E&2.94&21.46& 8.45&23.25&-&-& 5.4&1.0&21.89&0.37&16.18&0.07&3.45&0.17&0.39
\\
\noalign{\smallskip}
VC0951&B&M&N&E&6.63&22.34&25.85&23.24&-&-&21.1&1.6&22.59&0.13&14.46&0.07&4.00&0.08&0.26    	 
\\
VC0951&H&M&N&E&3.99&19.96&18.33&19.55&-&-&17.2&1.3&19.37&0.14&11.60&0.06&3.25&0.01&0.04    	 
\\
\noalign{\smallskip}
VC0965&B&M&B&E&0.92&21.65&21.67&23.42&-&-&23.0&1.4&23.72&0.10&15.54&0.06&3.45&0.03&0.01    	 
\\
VC0965&H&M&B&E&1.83&19.76&29.34&20.46&-&-&30.1&7.9&20.55&0.26&11.66&0.31&2.70&1.02&0.01    	 
\\
\noalign{\smallskip}
VC0975&H&M&N&E&6.79&20.46&49.74&21.28&-&-&48.5&3.3&21.20&0.09&11.06&0.09&3.03&0.14&0.04    	 
\\
\noalign{\smallskip}
VC1010&B&M&Y&E&8.55&21.89&24.85&22.53&-&-&21.9&1.4&22.09&0.10&13.89&0.07&3.13&0.05&0.18
\\
\noalign{\smallskip}
VC1011&H&T&N&E&-&-&34.35&20.53&20.79&19.89&23.5&1.2&20.26&0.08&12.30&0.06&2.46&0.06&0.44   	 
\\
\noalign{\smallskip}
VC1036&B&M&Y&E&7.82&21.22&31.01&22.90&-&-&21.6&1.6&21.80&0.12&14.06&0.06&4.33&0.05&0.28    	 
\\
VC1036&H&M&Y&E&7.98&17.55&30.02&19.48&-&-&18.5&1.5&17.99&0.14&10.65&0.06&4.41&0.08&0.36    	 
\\
\noalign{\smallskip}
VC1047&H&D&N&-&23.93&17.40&-&-&-&-&25.9&1.7&17.56&0.12&8.71&0.06&7.42&0.21&1.00            	 
\\
\noalign{\smallskip}
VC1073&B&M&N&E&5.14&21.76&26.09&23.32&-&-&24.6&1.5&23.05&0.10&14.48&0.06&3.76&0.01&0.20    	 
\\
VC1073&H&M&N&E&2.77&17.87&12.30&18.58&-&-&10.8&1.1&18.15&0.20&11.41&0.06&3.79&0.06&0.11    	 
\\
\noalign{\smallskip}
VC1078&B&E&N&-&-&-&12.73&22.90&-&-&12.2&1.1&22.87&0.17&16.01&0.06&3.20&0.03&0.00           	 
\\
VC1078&H&E&N&-&-&-&11.44&20.15&-&-&11.7&1.1&20.20&0.17&13.37&0.06&2.74&0.01&0.00           	 
\\
\noalign{\smallskip}
VC1122&B&M&Y&E&9.05&22.08&28.88&23.71&-&-&16.6&1.4&22.04&0.15&15.00&0.06&4.13&0.11&0.43    	 
\\
VC1122&H&E&Y&-&12.37&18.67&-&-&-&-&11.8&1.0&18.75&0.16&11.99&0.06&3.48&0.03&0.00           	 
\\
\noalign{\smallskip}
VC1173&B&E&N&-&-&-&10.61&22.92&-&-&11.3&1.1&23.05&0.17&16.43&0.06&2.76&0.01&0.00           	 
\\
VC1173&H&E&N&-&-&-&11.77&19.79&-&-&12.1&1.1&19.89&0.17&13.20&0.06&2.84&0.01&0.00           	 
\\
\noalign{\smallskip}
VC1183&H&M&Y&E&4.48&17.89&23.71&19.65&-&-&15.3&1.6&18.44&0.19&11.14&0.06&5.54&0.06&0.22    	 
\\
\noalign{\smallskip}
VC1254&B&M&B&E&1.44&19.92&13.02&23.32&-&-&12.3&1.2&23.13&0.17&15.73&0.06&3.25&0.01&0.22    	 
\\
VC1254&H&M&B&E&0.93&16.97&13.02&19.47&-&-&12.2&1.4&19.33&0.23&12.07&0.06&3.39&0.05&0.05    	 
\\
\noalign{\smallskip}
VC1308&B&M&Y&E&6.87&22.49&19.25&24.11&-&-&11.1&1.2&22.31&0.20&15.75&0.07&3.77&0.04&0.44    	 
\\
VC1308&H&E&Y&E&7.84&18.84&-&-&-&-&7.5&1.0&18.78&0.26&12.73&0.06&3.13&0.10&0.00             	 
\\
\noalign{\smallskip}
VC1348&B&M&B&E&1.79&21.73&9.58&22.93&-&-&8.4&1.1&22.57&0.24&16.01&0.07&3.35&0.07&0.09      	 
\\
VC1348&H&M&B&E&1.19&17.77&8.26&18.99&-&-&7.7&1.0&18.78&0.25&12.42&0.06&3.19&0.10&0.06      	 
\\
\noalign{\smallskip}
VC1386&B&M&Y&E&16.57&23.73&52.87&25.27&-&-&34.7&4.0&24.00&0.17&14.82&0.10&4.15&0.34&0.35   	 
\\
VC1386&H&E&Y&-&19.79&20.17&-&-&-&-&20.90&1.2&20.35&0.10&12.07&0.06&3.08&0.01&0.00           	 
\\
\noalign{\smallskip}
VC1392&H&T&Y&E&-&-&44.82&20.31&16.76&19.22&22.8&1.2&19.99&0.09&12.04&0.07&2.01&0.01&0.41   	 
\\
\noalign{\smallskip}
VC1453&B&M&N&E&4.79&21.82&20.40&22.97&-&-&19.1&1.4&22.69&0.12&14.46&0.07&3.62&0.05&0.18    	 
\\
VC1453&H&M&N&E&3.07&18.28&10.86&18.60&-&-&10.1&1.1&18.31&0.19&11.58&0.07&3.15&0.07&0.10    	 
\\
\noalign{\smallskip}
VC1491&B&M&N&E&7.78&22.23&28.10&24.77&-&-&13.3&1.3&22.46&0.17&15.26&0.07&4.48&0.11&0.41    	 
\\
VC1491&H&M&N&E&7.23&18.48&36.41&21.70&-&-&15.1&5.1&19.13&0.55&11.64&0.20&5.75&1.77&0.46    	 
\\
\noalign{\smallskip}
VC1499&B&M&N&E&4.92&20.51&12.67&23.59&-&-&7.4&1.1&21.26&0.26&15.18&0.07&3.24&0.06&0.64     	 
\\
VC1499&H&E&N&-&7.21&18.39&-&-&-&-&7.7&1.0&18.47&0.25&12.57&0.07&2.74&0.04&0.00             	 
\\
\noalign{\smallskip}
VC1514&H&E&?&-&-&-&24.37&19.99&-&-&28.1&1.5&20.55&0.09&12.19&0.06&2.88&0.03&0.00           	 
\\
\noalign{\smallskip}
VC1528&H&M&N&E&2.74&17.01&10.06&18.60&-&-&7.9&1.1&17.83&0.25&11.45&0.07&3.56&0.14&0.18     	 
\\
\noalign{\smallskip}
VC1549&H&M&Y&E&4.61&18.27&14.25&19.26&-&-&11.6&1.1&18.61&0.18&11.44&0.06&3.47&0.01&0.20    	 
\\
\noalign{\smallskip}
VC1684&B&M&B&E&1.95&22.02&17.39&22.58&-&-&20.3&1.5&22.74&0.13&15.46&0.06&3.19&0.07&0.03    	 
\\
\noalign{\smallskip}
VC1684&H&M&B&E&2.08&21.08&19.76&19.82&-&-&20.6&1.1&19.89&0.09&12.64&0.06&2.68&0.00&0.01    	 
\\
\noalign{\smallskip}
VC1695&H&M&N&E&4.95&18.28&19.00&20.29&-&-&13.7&1.2&19.25&0.16&11.74&0.07&3.65&0.02&0.21    	 
\\
\noalign{\smallskip}
VC1834&H&T&Y&E&-&-&17.21&17.57&12.16&16.46&16.7&1.2&17.67&0.12&10.05&0.07&3.12&0.01&0.77   	 
\\
\noalign{\smallskip}
VC1895&H&M&N&E&7.13&18.94&19.89&20.01&-&-&13.8&1.2&18.86&0.14&12.11&0.08&3.89&0.20&0.24    	 
\\
\noalign{\smallskip}
VC1910&B&M&N&E&4.55&21.92&14.20&22.32&-&-&13.2&1.1&22.03&0.15&14.56&0.07&3.07&0.01&0.12
\\
\noalign{\smallskip}
VC1947&H&E&Y&-&-&-&8.15&17.54&-&-&9.2&1.1&17.69&0.22&11.13&0.07&3.05&0.05&0.00             	 
\\
\noalign{\smallskip}
VC2042&H&E&Y&-&-&-&21.20&20.63&-&-&22.2&1.1&20.79&0.09&12.13&0.06&2.73&0.01&0.00           	 
\\
\noalign{\smallskip}
VC2050&H&E&N&-&-&-&11.64&19.00&-&-&13.1&1.1&19.28&0.15&12.37&0.06&2.85&0.02&0.00           	 
\\
\noalign{\smallskip}
2MASSX&H&D&N&-&4.19&17.01&-&-&-&-&5.2&1.1&17.47&0.40&12.10&0.06&5.90&0.02&1.00             	 
\\
\noalign{\smallskip}
CCC045&H&D&N&-&4.30&16.13&-&-&-&-&4.5&1.1&16.29&0.46&11.10&0.06&6.23&0.71&1.00             	 
\\
\noalign{\smallskip}
CCC059&H&M&N&E&0.57&18.17&4.27&19.97&-&-&4.1&1.1&19.81&0.47&15.00&0.10&3.38&0.21&0.10      	 
\\
\noalign{\smallskip}
CCC094&H&M&N&E&2.76&18.59&12.75&19.95&-&-&11.3&1.1&19.60&0.18&12.50&0.06&3.65&0.07&0.14    	 
\\
\noalign{\smallskip}
CCC095&H&M&N&D&2.62&14.69&25.07&19.33&-&-&11.6&1.2&17.02&0.19&10.16&0.06&8.42&0.37&1.00    	 
\\
\noalign{\smallskip}
CCC096&H&M&Y&E&8.20&17.37&17.97&18.67&-&-&12.0&1.1&17.14&0.17&10.54&0.07&3.65&0.03&0.46    	 
\\
\noalign{\smallskip}
CCC104&H&E&N&-&-&-&3.80&19.59&-&-&3.9&1.0&19.61&0.53&14.78&0.06&2.81&0.17&0.00             	 
\\
\noalign{\smallskip}
CCC113&H&M&N&E&1.44&17.61&4.36&18.73&-&-&3.4&1.0&17.93&0.60&13.42&0.07&3.35&0.35&0.25      	 
\\
\noalign{\smallskip}
CCC119&H&M&N&E&2.68&15.71&11.48&18.45&-&-&5.8&1.1&16.44&0.34&10.75&0.06&3.95&0.16&0.46     	 
\\
\noalign{\smallskip}
CCC122&H&M&N&E&2.49&14.43&20.01&17.91&-&-&8.1&1.2&15.38&0.28&9.59&0.07&7.24&0.38&0.34      	 
\\
\noalign{\smallskip}
CCC125&H&E&N&-&-&-&7.59&19.72&-&-&8.0&1.0&19.85&0.25&13.56&0.06&2.62&0.04&0.00             	 
\\
\noalign{\smallskip}
CCC136&H&M&N&E&0.88&16.07&3.89&17.84&-&-&3.0&1.0&17.04&0.70&12.79&0.07&3.84&0.58&0.21      	 
\\
\noalign{\smallskip}
CCC142&H&E&N&-&-&-&3.81&19.36&-&-&3.5&1.0&19.16&0.58&14.62&0.06&2.18&0.02&0.00             	 
\\
\noalign{\smallskip}
CCC150&H&E&N&-&-&-&4.21&20.01&-&-&4.2&1.0&20.05&0.48&15.16&0.06&3.11&0.23&0.00             	 
\\
\noalign{\smallskip}
CCC153&H&T&Y&E&-&-&4.80&19.12&3.06&18.05&3.8&1.0&18.88&0.52&14.53&0.07&2.63&0.12&0.67      	 
\\
\noalign{\smallskip}
CCC157&H&E&N&-&-&-&5.87&20.44&-&-&5.8&1.0&20.39&0.34&15.17&0.07&2.70&0.05&0.00             	 
\\
\noalign{\smallskip}
CCC205&H&M&N&E&1.51&16.02&7.90&18.67&-&-&4.6&1.1&17.08&0.46&11.96&0.06&5.28&0.72&0.32      	 
\\
\noalign{\smallskip}
CCC216&H&E&N&-&-&-&7.68&20.83&-&-&7.0&1.1&20.57&0.29&14.77&0.06&2.76&0.06&0.00             	 
\\
\noalign{\smallskip}
CCC222&H&M&N&E&1.41&18.01&8.90&18.88&-&-&9.3&1.1&18.84&0.22&12.20&0.06&3.22&0.04&0.06      	 
\\
\noalign{\smallskip}
CCC226&H&T&?&D&24.03&17.49&-&-&10.49&15.64&9.1&1.2&15.89&0.25&10.03&0.07&5.14&0.19&0.90    	 
\\
\noalign{\smallskip}
CEG050&H&M&N&E&1.18&19.13&3.38&19.27&-&-&3.1&1.0&18.96&0.67&14.63&0.07&3.10&0.27&0.11      	 
\\
\noalign{\smallskip}
97073&B&T&?&E&-&-&15.68&22.74&5.67&19.92&10.3&1.1&22.46&0.18&15.66&0.07&1.98&0.02&0.68     	 
\\
97073&H&E&?&-&-&-&10.04&19.87&-&-&10.4&1.0&19.94&0.19&13.19&0.06&2.74&0.02&0.00                 	 
\\
\noalign{\smallskip}
97087&B&E&?&-&-&-&15.19&20.05&-&-&26.1&1.2&21.37&0.07&14.31&0.07&1.87&0.01&0.00            	 
\\
97087&H&E&?&-&-&-&12.59&16.72&-&-&14.4&1.2&16.98&0.15&11.01&0.07&3.45&-&0.00  	 
\\
\noalign{\smallskip}
IZW018&H&E&?&-&-&-&2.02&19.39&-&-&2.0&1.0&19.40&0.99&16.06&0.10&2.84&-&0.00             \\
\end{longtable}
\end{landscape}
\normalsize

\end{document}